\documentclass[conference,10pt]{IEEEtran}
\usepackage{epsfig,endnotes,url}
\usepackage{xspace}
\usepackage{xcolor,graphicx}
\usepackage{multicol}
\usepackage{subcaption}

\begin{document}

%don't want date printed
\date{}

\title{Virtually the Same:\\Comparing Physical and Virtual Testbeds}

\author{\IEEEauthorblockN{Jonathan Crussell, Thomas M Kroeger, Aaron Brown, Cynthia Phillips}
\IEEEauthorblockA{\textit{Sandia National Laboratories} \\
\{jcrusse, tmkroeg, aarbrow, caphill\}@sandia.gov}
}

\maketitle

\section*{Abstract}

Network designers, planners, and security professionals increasingly rely on
large-scale testbeds based on virtualization to emulate networks and make
decisions about real-world deployments. However, there has been limited
research on how well these virtual testbeds match their physical counterparts.
Specifically, does the virtualization that these testbeds depend on actually
capture real-world behaviors sufficiently well to support decisions?

As a first step, we perform simple experiments on both physical and virtual
testbeds to begin to understand where and how the testbeds differ. We set up a
web service on one host and run ApacheBench against this service from a
different host, instrumenting each system during these tests.

We define an initial repeatable methodology (algorithm) to quantitatively
compare physical and virtual testbeds. Specifically we compare the testbeds at
three levels of abstraction: application, operating system (OS) and network.
For the application level, we use the ApacheBench results. For OS behavior, we
compare patterns of system call orderings using Markov chains. This provides a
unique visual representation of the workload and OS behavior in our testbeds.
We also drill down into read-system-call behaviors and show how at one level
both systems are deterministic and identical, but as we move up in abstractions
that consistency declines. Finally, we use packet captures to compare network
behaviors and performance. We reconstruct flows and compare per-flow and
per-experiment statistics.

From these comparisons, we find that the behavior of the workload in the
testbeds is similar but that the underlying processes to support it do vary.
The low-level network behavior can vary quite widely in packetization depending
on the virtual network driver. While these differences can be important, and
knowing about them will help experiment designers, the core application and OS
behaviors still represent similar processes.

\section{Introduction}
\label{sec:intro}

Network operators and designers rely heavily on testbeds to verify configuration
changes, validate new designs, and troubleshoot existing networks without
interrupting production services. In some cases, security engineers use testbeds
to better understand security incidents, do post-event forensics, and even
explore how various countermeasures would perform. Physical testbeds are
expensive to build and maintain so virtual testbeds are used as a cost-effective
alternative. By running the same OS and software, virtualization offers a higher
fidelity than discrete-event simulations. But how do virtualization artifacts
affect these testbeds and the experiments they host? How should experimenters
account for subtle differences between virtual and physical testbeds?

To date, there has been limited study of the effects of virtualization on
virtual testbeds. Instead, we rely on ad hoc assessments of the experiments and
results by subject-matter experts to determine whether they make logical sense.
We know that virtualization causes higher network latency and lowers
throughput~\cite{menon2005diagnosing,rizzo2013speeding,wang2010impact,whiteaker2011explaining}
for individual virtual machines (VMs). Wang et al.~\cite{wang2010impact} even
showed that for small bursts, buffering can cause VMs to receive data at rates
that exceed the underlying network.

Our experimental design is informed by our goal: to measure, document, and
understand differences between physical and virtual testbeds. We run a
representative workload, ApacheBench~\cite{apachebench} querying a simple web
server. Our experiments vary testbed parameters such as the network driver and
workload parameters such as payload size. During these experiments, we measure
the systems at three layers of abstraction: application, operating system, and
underlying network.

We find notable differences in the system and network-level interactions. For
example, when receiving a 1MB payload, all testbeds read a total of 1MB of
data. However, they use vastly different numbers of \emph{read} system calls
due to differences in segmentation offloading. Additionally, they transmit
differing numbers of packets. These differences in underlying system behaviors
are important and have the potential to affect experimental results.

When one considers that computers are fundamentally state machines that cycle
through instructions deterministically, one might expect physical and virtual
testbeds to have (nearly) identical behaviors. This is particularly true for our
workload, which has no inter-request dependencies. One way to model an
application's behavior is through its sequence of system calls. Since our
workload consists of many repetitions of the same transaction, we expect the
system call traces to contain many repeated sequences. To make the sequences
simpler to work with and to normalize the repeated behaviors across sequences of
different lengths (caused by different testbeds completing different numbers of
transactions), we transform the sequences into Markov Chains as described in
Section~\ref{sec:background}. In Section~\ref{sec:markov}, we show how the
Markov chain for the physical and virtual testbeds can create almost identical
graph topologies representing system behaviors.

Even though one might expect a deterministic workload, there is variation even
when running the same workload on the same physical testbed. So, we must expect
some differences between virtual and physical testbeds. Our goal is not to prove
whether virtual testbeds are good or bad. It is to understand where and how they
differ from physical testbeds so those who use virtual testbeds will be better
able to plan their experiments and interpret the results.

\section{Background}
\label{sec:background}

Network simulation tools like NS-2~\cite{issariyakul2011introduction} and
OPNET~\cite{chang1999network} create highly detailed network models to simulate
a network's behavior under many conditions. To ensure simulation reliability,
the tools contain simulated versions of all aspects of a network, including
endpoints, routers, and switches. These network models follow a set of known
behaviors exactly. However, this reliable repetition comes at the cost of
decreased accuracy. The simulated network elements model actual components
that, in the real world, may display behaviors different from the tool's
built-in components.

Network emulation improves the accuracy of these network models by introducing
actual network components. But it is more difficult to scale simulated testbeds
to emulate large-scale network environments. Virtualization, utilizing multiple
VMs on the same physical machine, permits testbeds to effectively emulate
networks containing thousands or even millions of
endpoints~\cite{minnich:2010eurosys}.

Sandia National Laboratories has been researching, developing, and applying
large-scale emulations using virtualization as testbeds for over a
decade~\cite{minnich:2010eurosys}. Sandia has developed several supporting
toolsets including minimega~\cite{minimega}.

Several other papers have presented testbed orchestration
platforms~\cite{benzel2011science,Chun2003Planetlab,lantz2010network,ricci2014Cloudlab}.
While our experiments could be run on any of these platforms and most modern
hardware, for this paper we focus on the different between physical and virtual
on one specific cluster using a single orchestration tool. This allows us to
experiment with different parameters within the virtual machine itself (e.g.
different network drivers).  Future work will expand across different testbed
platforms.

Multiple researchers expose differences in latency and throughput in
virtualized networking
devices~\cite{menon2005diagnosing,rizzo2013speeding,wang2010impact,whiteaker2011explaining}.
Virtualization can also cause different network behaviors, especially around
TCP and congestion
control~\cite{Cheng:2016TransactionNetworking,Gamage:2013:TransCompSys,He:2016Sigcomm}.
These papers typically present new approaches for performance improvement. We
focus on understanding the differences and how they affect the accuracy of
emulations used to understand larger-scale phenomena.

Sequences of system calls can capture the expected behavior of applications in
areas such as intrusion detection~\cite{forrest1996sense,hofmeyr1998intrusion}
and filesystem optimization~\cite{kroeger1999case}. Previous
research~\cite{warrender1999detecting} used system-call Markov chains for
intrusion detection. To the best of our knowledge, this is the first paper to
apply such Markov chains to compare virtual to physical testbeds.

A {\em Markov chain} is a graph where nodes represent states and edges
represent transitions between states. For a sequence of system calls, we
empirically create a graph where a node is a system call and two nodes, $X$ and
$Y$, are connected by a directed edge (arc) from $X$ to $Y$ i.f.f.\@ the system
call $Y$ appears immediately after system call $X$ at least once. Each arc has
a weight corresponding to the probability that $Y$ follows $X$. The weight of
arc $(X,Y)$ is the percentage of times $Y$ is immediately followed by $X$ in
the sequence.

\begin{figure}
    \centering
    \includegraphics[width=0.9\columnwidth]{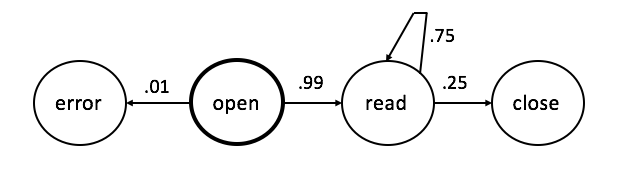}
    \caption{An example of a Markov chain for system calls seen in searching a
    file line by line.}
    \label{fig:Markov-eg}
\end{figure}

Figure~\ref{fig:Markov-eg} shows a simple Markov chain representing a single
user search for an item in a file. The user first tries to open the file. This
is the start state, shown with a heavy node boundary. With probability $.01$,
there is an error in opening the file, where the new state is ``error,'' and
the process ends.  With probability $.99$, the open succeeds and the user code
reads the first line.  For each line, the user finds what he seeks with
probability $.25$. If so, the user closes the file and the search ends.
Otherwise, with probability $0.75$ the user continues the search by reading the
next line, and the state (current node) does not change.

We can extend this definition to include more of the recent history of system
calls by having states represent a set of $N$ consecutive system calls. If a
node has an ordered set of calls (label) $c_1,c_2, \ldots, c_N$, then the next
state, represented by neighbors in the graph, has the form
$c_2, c_3, \ldots, c_N, c_{\mbox{new}}$, where $c_{\mbox{new}}$ is the next
system call seen after $c_N$. As $N$ increases, the Markov chain becomes
specific to the sequence of system calls that generated it, diminishing the
chances of it matching system-call sequences from another process.

We now formalize the ``similar/close'' concepts we have used so far. We say
that two experimentally measured quantities are ``close'' or ``similar'' if
their means differ by no more than $10$\%. This is an arbitrary value for
quantitative comparisons in this paper.  For a particular application, a
subject-matter expert may need to set this comparison point.

We compute $95$\% confidence intervals, assuming a standard normal distribution
of sample means. Assuming our samples are representative of a random sample of
all such experiments, then if we were to repeat the experiment with this size
random sample an infinite number of times, the sample mean will be within the
confidence interval range $95$\% of the time. Overlapping confidence intervals
is a stronger definition.

\section{Methodology}
\label{sec:methodology}

In this section, we detail a methodology to compare workloads across testbeds,
whether physical or virtual. Although we consider a single workload in this
paper, we ultimately must consider many different network-based workloads using
this basic methodology.

The simplest way to compare testbeds is with metrics from the workload itself,
such as how much work completes in a fixed amount of time. However, the
virtualization overhead almost always leads to physical testbeds outperforming
the virtual ones. Therefore, we also take measurements at other levels of
abstraction -- interactions between the workload and the OS. Specifically, we
measure system calls, context switches, and block IO and normalize by work
completed to determine the interactions per unit work. This allows direct
comparisons between testbeds. Finally, we compare our network-based workload at
the network layer. We capture traffic to compare properties of the underlying
packets and flows (for TCP-based workloads). These different layers of
abstraction allow us to compare behaviors, not just performance.

If there were just one physical and one virtual testbed, our comparison would
be fairly straightforward -- we would apply the above methodology to both
across several workloads and draw our conclusions. Unfortunately, there are
many physical and virtual testbeds. These testbeds vary in machine resources
(e.g.\@ CPU and memory), network bandwidth (e.g.\@ 100Mbps, 1Gbps), network
interface and driver, OSes, etc. To draw broader conclusions about the
differences between physical and virtual testbeds, we must perform tests with
many parameters to determine if there are any generalizable trends. We do not
claim to fully map this space but present an initial mapping that may help us
to understand the landscape.

Virtualization allows for multiple VMs on the same machine meaning that virtual
testbeds can be many times larger than the underlying physical machines, a
practice called oversubscription, which induces
contention~\cite{minnich:2010eurosys}. As a first step, we assume that our
systems are not resource constrained. We believe a strong understanding of
normal/non-resource-constrained behavior should come before studying edge cases
introduced by oversubscription.

\section{Experimental Design}
\label{sec:experiment}

We detail a concrete application of our methodology to a simple HTTP workload.
We describe the testbeds, workload, and instrumentation. We repeated each
experiment ten times.

\subsection{Virtual Testbed Tool: minimega}

We use minimega~\cite{minimega} to manage, deploy, and monitor our virtual
testbeds. It is the product of over a decade of research and development at
Sandia National Laboratories. minimega orchestrates Kernel-based Virtual
Machines (KVM~\cite{kivity2007kvm}) to run unmodified OSes such as Windows,
Linux, and Android, that represent real machines in networks of interest.
minimega is bundled with other tools to form the open-source minimega toolset,
which supports our Emulytics program.

When configuring VMs, the user decides the number of network interfaces,
network connectivity, and network drivers. minimega uses 802.1q VLAN tagging
via Open vSwitch~\cite{openvswitch} to support arbitrary network topologies.
KVM supports several network drivers with varying properties such as
\emph{e1000} and \emph{virtio}. \emph{e1000} represents a \textit{real}
\emph{e1000} network interface card (NIC) while \emph{virtio} is a
\textit{paravirtualized} NIC that is implemented specifically for better
performance within a VM.

minimega has many other capabilities to support experiments. Its
command-and-control layer can execute commands on the VMs and push/pull files.
It supports packet captures for individual VMs. It can emulate networks with
different speeds based on Linux traffic control, ``tc.'' For example, minimega
can emulate a 1Gbps switch by rate limiting all VMs on a network to 1Gbps.
While minimega focuses on virtual testbeds, it can also deploy and run
experiments on physical testbeds, as we did for our physical experiments.

We use several other tools in the minimega toolset. \emph{protonuke} is a
simple traffic generator that supports several protocols and acts as either a
server or client. \emph{vmbetter} creates initial ramdisk images that we use
for physical hosts and VMs.

\subsection{Workload}

We use ApacheBench~\cite{apachebench}, the HTTP server benchmarking tool, to
repeatedly load the same page served by \emph{protonuke}, acting as a simple
HTTP server. Client and server run on separate (virtual) machines, connected by
a switch. This transaction, an HTTP GET and response, is at the core of almost
all of the more complex workloads we test. By keeping to this basic workload we
can focus on key system interactions.

We can vary the number of client threads, the number of requests or duration,
and HTTP response size. Our experiments use three response sizes: 500B, 1MB and
16MB. We run ApacheBench for 90 seconds, allowing it to complete as many
requests as possible (up to 500,000 requests). We use ten client threads,
except where noted.

\subsection{Virtual Machines}

There are many ways to instantiate VMs. We can vary hypervisors, network
drivers, CPU pinning, scheduling algorithms, resource allocations, host OS,
etc. We focus on the network driver, the glue between the guest OS and the host
OS. Specifically, we look at the differences between \emph{e1000}, which
emulates a real e1000 network interface and \emph{virtio}, which was designed
to make VMs more performant. We chose these two drivers because e1000 is the
default in minimega (and therefore, often unchanged) and because virtio is the
go-to driver to improve networking performance (at least within the Emulytics
community). We did not vary the other parameters -- we used KVM for the
hypervisor, 2GB of memory, and 8 vCPUs. For now, these parameters represent a
well-provisioned system able to support the workload.

We base our VMs off an initial ramdisk (initrd) from \emph{vmbetter}. This
allows us to create a minimal install containing few packages and a minimal
init program performing only critical functions. Specifically, the init process
initializes the filesystem, loads kernel modules, starts sshd, and runs
\emph{miniccc} (the agent with which minimega communicates). This minimizes
background-process interference. We do the same for our physical machines.

\subsection{Hardware}

We ran the physical tests on two identical nodes:

\begin{itemize}
    \itemsep 0em
    \item Dual Intel(R) Xeon(R) E5-2630L 0 @ 2.00GHz
    \item 24 Cores (6/socket, HT enabled), 125GB Memory
    \item Interfaces: 1Gbps (igb), 10Gbps (ixgbe)
\end{itemize}

We ran the VM tests on two identical nodes:

\begin{itemize}
    \itemsep 0em
    \item Dual Intel(R) Xeon(R) E5-2683 v4 @ 2.10GHz
    \item 64 Cores (16/socket, HT enabled), 503GB Memory
    \item Interfaces: 100Gbps (mlx5)
\end{itemize}

These nodes represent our older and newer Emulytics nodes which are well
provisioned to run many VMs. The older nodes have both 1Gbps and 10Gbps
interfaces, so we can perform tests with both link speeds. The newer nodes have
100Gbps interfaces. For each interface, we list the associated driver.

All physical and virtual machines use the same Linux Kernel version,
4.9.0-4-amd64, the latest available from Debian stretch when we built the
images.

\subsection{Instrumentation}
\label{sub:instrumentation}

In Section~\ref{sec:methodology}, we described the various levels of
instrumentation that we use to compare testbeds. Here, we describe the specific
instrumentation tools that we use. In Section~\ref{sec:overhead}, we perform
experiments to understand the overhead of instrumentation on the workload.

\emph{System Call Traces:} We collect system-wide system call traces using
\emph{sysdig}~\cite{sysdig}. These traces contain all calls from all programs
running on the physical or virtual machine.

\emph{Packet Captures (PCAPs):} For the physical machines, we capture traffic
on the machine itself using \emph{tcpdump}~\cite{tcpdump}. For the VMs, we
capture traffic from the host using minimega (which uses \emph{libpcap}) to
avoid the performance overhead of capturing within a VM. Since \emph{tcpdump}
also uses \emph{libpcap}, we do not expect the capture method to introduce any
significant differences. We use \emph{tcptrace}~\cite{tcptrace} which assembles
TCP packets from the PCAPs into flows and computes statistics, such as the
number of packets, and retransmits, on a per-flow basis.

\emph{Latency:} To measure queueing, we use \emph{owping}~\cite{owamp} to
measure the one-way latency between the client and server. We do not
synchronize the clocks so we cannot use the absolute latency values. Instead,
we use the jitter to infer how much the latency varied over the duration of the
experiment.

\section{Results}
\label{sec:results}

In this section, we present our initial results including experimental data and
metric comparisons between the physical and virtual testbeds. We analyze the
application-level metrics which confirm that the workload behaves as expected,
although at different rates. Then, we explore how the underlying interactions
with the OS and network vary between the testbeds.

\subsection{Application-level Metrics}

Here we examine the results from ApacheBench for physical, e1000, and virtio
drivers for both 1Gpbs and 10Gbps systems. We find an anomaly in e1000 tests
but otherwise the results seem quite comparable. We then explore issues with
our e1000 results and correlate the anomaly to a known bug.

\begin{table}[ht]
    \centering
    \begin{tabular}{|r|r|r|r|}
        \hline
        Size  & Physical & e1000 & Virtio \\
        \hline
        500B  &  14420  $\pm$  74.3   &  6476  $\pm$  707  &  13590  $\pm$  139\\
        1MB   &  112    $\pm$  0.012  &  113   $\pm$  0.12  &  113    $\pm$  0.006   \\
        16MB  &  6.97   $\pm$  0.006  &  7.05  $\pm$  0.006  &  7.09   $\pm$  0.032   \\
        \hline
    \end{tabular}
    \caption{Mean requests per second and confidence intervals for ApacheBench
    runs for 1Gbps tests.}
    \label{tab:ab_rps_1g}
\end{table}

Table~\ref{tab:ab_rps_1g} shows the confidence intervals for mean requests per
second for the 1Gbps tests. With the exception of e1000 for small workloads,
which we explore later in this section, all systems have similar results. The
network driver can have a significant impact on much higher-level application
behavior and how well an emulation resembles the physical world. Moreover, the
selection of workload size also impacts how representative our results are.

\begin{table}[ht]
    \centering
    \begin{tabular}{|r|r|r|r|}
        \hline
        Size  & Physical & e1000 & Virtio \\
        \hline
        500B  &  13080  $\pm$  101   &  6734  $\pm$  867  &  13631  $\pm$  139   \\
        1MB   &  638    $\pm$  2.4  &  144   $\pm$  20.6  &  590    $\pm$  35.7  \\
        16MB  &  50.0   $\pm$  0.316  &  13.2  $\pm$  0.955  &  44.5   $\pm$  2.78  \\
        \hline
    \end{tabular}
    \caption{Mean requests per second and confidence intervals for ApacheBench
    runs for across 10Gbps tests.}
    \label{tab:ab_rps_10g}
\end{table}

Table~\ref{tab:ab_rps_10g} shows these same test with an emulated 10Gbps
network. Here the differences for e1000 are more pronounced. Once again, we see
that the behaviors for physical and virtio are similar. For small payload
sizes, virtio actually outperforms the physical system. We speculate and have
seen anecdotal evidence that this performance difference may be an artifact of
the way our bandwidth-limiting tool emulates a 10Gbps link. At this higher
rate, the physical testbed is noticeably more consistent in its behavior than
the virtual ones.

\subsubsection{Instrumentation Overhead}
\label{sec:overhead}

To quantify the overhead of our instrumentation, we ran a set of tests with no
instrumentation. Table~\ref{tab:no_inst} shows our ApacheBench results from
this test. The most significant impact of our instrumentation is on the
physical testbed where there is a 13\% drop in performance. We believe this is
from the overhead of running tcpdump on the physical testbed. In the virtual
testbeds, virtio drops by 5\% while e1000 improves. We suspect the latter is
due to high variability in the e1000 performance.

\begin{table}[ht]
    \centering
    \begin{tabular}{|r|r|r|r|}
        \hline
        Size  & Physical & e1000 & Virtio \\
        \hline
        \hline
        \multicolumn{4}{|c|}{1Gbps}\\
        \hline
        500B  &  16459  $\pm$  57    &  5740  $\pm$  559   &  14373  $\pm$  182.8   \\
        16MB  &  6.98   $\pm$  0.0062  &  7.05  $\pm$  0.0062  &  7.09   $\pm$  0.012  \\
        \hline
        \hline
        \multicolumn{4}{|c|}{10Gbps}\\
        \hline
        500B  &  15013  $\pm$  104   &  5474  $\pm$  374  &  14431  $\pm$  227  \\
        16MB  &  60.1   $\pm$  0.39  &  11.7  $\pm$  1.61  &  42     $\pm$  1.74  \\
        \hline
    \end{tabular}
    \caption{Mean request per second and confidence intervals for ApacheBench
    runs without instrumentation.}
    \label{tab:no_inst}
\end{table}

\subsubsection{Exploring the Outlier, e1000}

To understand why our e1000 testbed behaves so differently, we turned to the
PCAPs that we collect. We observed that numerous experiments had an errant
behavior that would delay connections for some multiple of 13 seconds. Upon a
more detailed examination, we saw that data had been sent and acknowledged but
the server was still behaving as if it had not been acknowledged (for all
connections). Then, after a multiple of 13 seconds the server returned to
normal behavior. From the kernel logs for the VM, we found that the kernel
reset the network adapter multiple times after detecting a transmit queue
timeout. Network adapter resets are commonly observed behavior when a bug in
the underlying NIC driver is encountered. We plan to discard experiments that
trigger this bug in the future.

\subsection{OS-Level Metrics}
\label{sec:os_level}

We now investigate differences in how ApacheBench interacts with the OS as it
makes requests. We focus on data read and read system calls as these are a key
parts of the workload.  System calls present a middle ground in abstraction
between the request per second and the bytes per request. We normalize using
the number of completed requests, to compare across testbeds completing vastly
different numbers of requests.

\begin{figure}
    \centering
    \begin{subfigure}[b]{0.23\textwidth}
        \includegraphics[width=\textwidth]{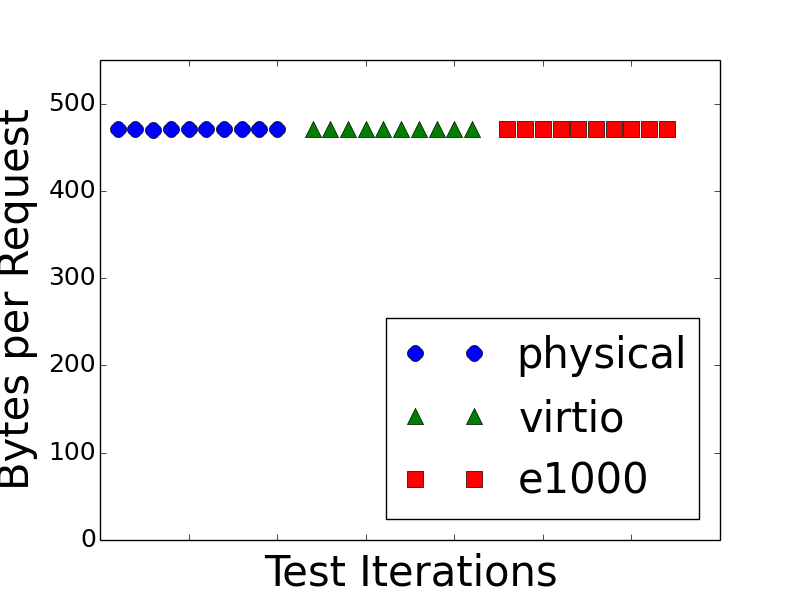}
        \caption{Bytes per Request}
        \label{fig:bytes_p_req}
    \end{subfigure}
    ~ %add desired spacing between images, e. g. ~, \quad, \qquad, \hfill etc.
    \begin{subfigure}[b]{0.23\textwidth}
      \includegraphics[width=\textwidth]{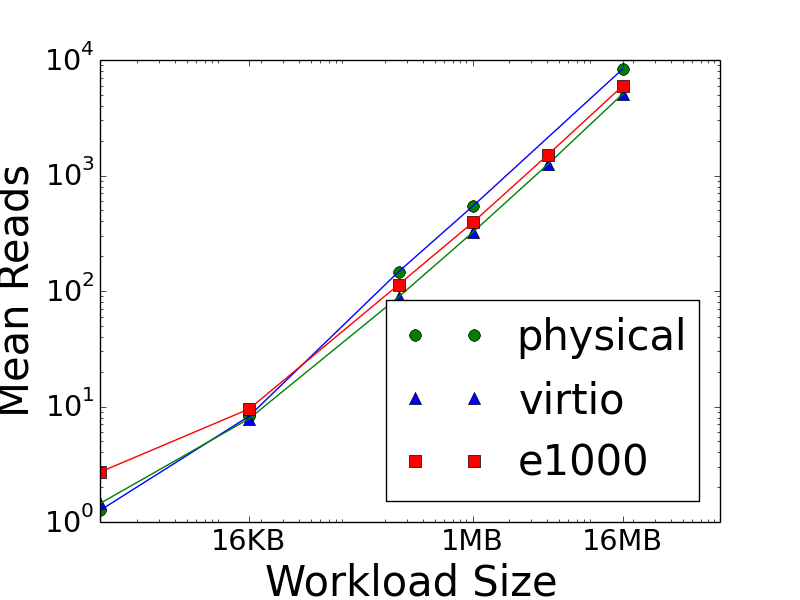}
      \caption{Mean read calls}
        \label{fig:reads_per_ab}
    \end{subfigure}
    \caption{(a) Bytes per Request for 500B workloads over 1Gpbs network. (b)
    Mean number of read system calls needed to complete a request across
    varying workload sizes.}
    \label{fig:read_figs}
\end{figure}

We look at a low-level metric that should be strongly consistent: bytes sent
per request completed. Figure~\ref{fig:bytes_p_req} plots the number of bytes sent
by the server per request completed, including retransmits. While it may seem
trivial to show that these numbers are consistent, this measure is independent
of timing and network rates so it provides a basic ability to show that the
underlying process is consistent and deterministic. This plot shows a clearly
consistent behavior from all testbeds.

Our underlying thesis for Emulytics is that since we run the same software, we
should get the same behavior. Since a thread is a deterministic machine that
works through the same lock steps in the same way for the same input, we expect
somewhere in our experiments to see this high level of consistency across all
tests.

At a higher level of abstraction, but still within the OS layer, we consider
number of reads per request. Table~\ref{tab:reads_per_request_1g} shows the
mean number of reads per request for the 1Gbps test. As expected, the number of
reads increases as the payload grows. In all but the small payload workload,
the physical testbed performs more reads than the virtual ones. In all cases,
virtio requires fewer reads than e1000.

\begin{table}[ht]
	\centering
	\begin{tabular}{|r|r|r|r|}
		\hline
		Size & Physical & e1000 & Virtio \\
		\hline
        500B  &  1.34     &  2.95     &  1.75     \\
        1MB   &  547.44   &  395.02   &  325.27   \\
        16MB  &  8345.82  &  5954.81  &  5072.02  \\
		\hline
	\end{tabular}
	\caption{Mean number of read system calls per request for 1Gbps test.}
	\label{tab:reads_per_request_1g}
\end{table}

Table~\ref{tab:reads_per_request_10g} shows the mean number of reads per
request for the 10Gbps test. Unlike in Table~\ref{tab:reads_per_request_1g},
the physical testbed has fewer reads per request than the virtual testbeds in
all but one case. This could be a result of the physical driver being different
for the 1Gbps and 10Gbps test which use \emph{igb} and \emph{ixgbe},
respectively.

\begin{table}[ht]
    \centering
    \begin{tabular}{|r|r|r|r|}
        \hline
        Size & Physical & e1000 & Virtio \\
        \hline
        500B  &  1.72     &  3.02     &  1.69     \\
        1MB   &  260.15   &  341.84   &  275.80   \\
        16MB  &  4098.74  &  4687.41  &  4102.61  \\
        \hline
    \end{tabular}
    \caption{Mean number of read system calls per request for 10Gbps test.}
    \label{tab:reads_per_request_10g}
\end{table}

Finally, we look more broadly at the number of read system calls across a wider
range of workload sizes. Figure~\ref{fig:reads_per_ab} shows the mean number of
read calls normalized over the mean requests per second from ApacheBench. Here
we begin to see the behavior of each process start to diverge.

\subsection{Network-Level Metrics}

From the traffic captures, we can gain a better understanding of how packets
generated by the various workloads propagate through the testbeds.
Table~\ref{tab:packets_per_request_1g} shows the mean packets per request for
1Gbps tests. As before, we see physical and virtio have similar behaviors,
though other than for the smallest response size, not quite at our $10$\%
similarity threshold. Again, e1000 stands out as different.

\begin{table}[ht]
    \centering
    \begin{tabular}{|r|r|r|r|}
        \hline
        Size & Physical & e1000 & Virtio \\
        \hline
        500B  &  5.00  $\pm$  0.08  &  5.00     $\pm$  0.10    &  5.00  $\pm$  0.12  \\
        1MB   &  67.7  $\pm$  2.19  &  105.7    $\pm$  9.04    &  77.3  $\pm$  10.2  \\
        16MB  &  834   $\pm$  46.3  &  1527  $\pm$  817.40  &  1087  $\pm$  706   \\
        \hline
    \end{tabular}
    \caption{Mean number of packets per request for 1Gbps test and standard deviation.}
    \label{tab:packets_per_request_1g}
\end{table}

Next, we used owamp data to measure the network jitter. This metric offers
insights into the network queuing behavior. If the network stack is filled with
many packets the jitter will indicate increased delays as the one-way latency
traffic queues while waiting for transmission over network links.
Table~\ref{tab:owamp_tab} shows how our jitter varied across our 1Gbps tests.
Here we see that e1000, with the exception of the 16MB workload, more closely
matches the end-to-end jitter characteristics seen in the physical testbed. The
virtio testbed provides consistently lower jitter which, while theoretically
beneficial, could mask problems that would occur in the real world if a given
workload was heavily influenced by jitter.

\begin{table}[ht]
	\centering
	\begin{tabular}{|r|r|r|r|}
		\hline
		Size & Physical & e1000 & Virtio \\
		\hline
        500B  &  0.26  $\pm$  0.03  &  0.25  $\pm$  0.074  &  0.10  $\pm$  0.00  \\
        1MB   &  0.29  $\pm$  0.019  &  0.24  $\pm$  0.03  &  0.16  $\pm$  0.03  \\
        16MB  &  0.28  $\pm$  0.025  &  0.34  $\pm$  0.043  &  0.16  $\pm$  0.03  \\
		\hline
	\end{tabular}
        \caption{Mean jitter measured across the network (in milliseconds) with
          confidence intervals for 1Gbps.}
	\label{tab:owamp_tab}
\end{table}

\subsection{System-Call Markov Chains}
\label{sec:markov}

We propose to use two analyses from the system-call traces to determine whether
the application behavior on the virtual testbeds are ``close enough'' to the
physical testbed. We examine a specific set of parameters for simplicity:
single thread, 1Gbps network, and 500B response. These analyses will need to be
repeated for every set of parameters to make broader comparisons between the
testbeds.

\begin{figure*}
    \centering
    \includegraphics[width=0.3\textwidth]{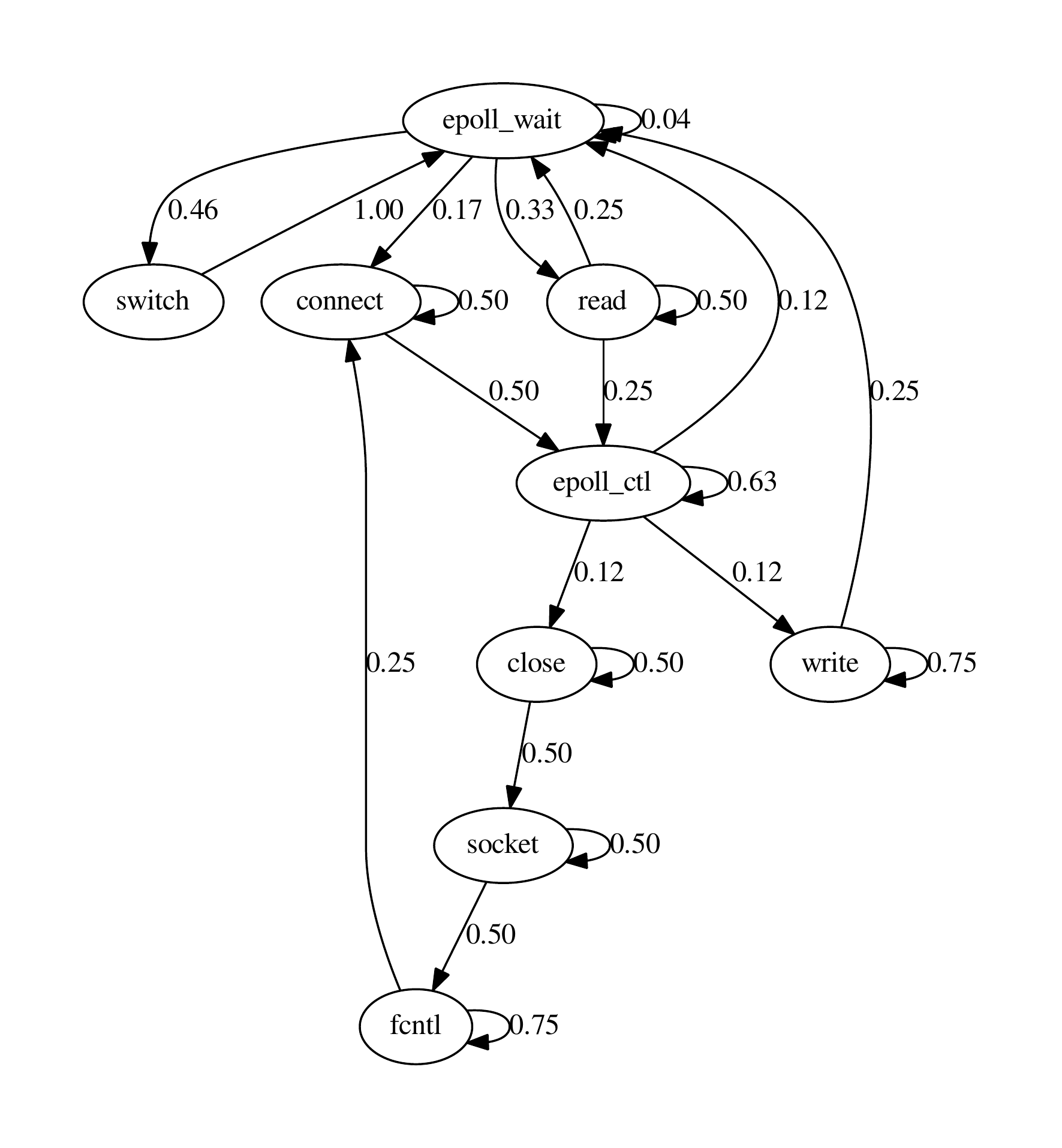}
    \includegraphics[width=0.3\textwidth]{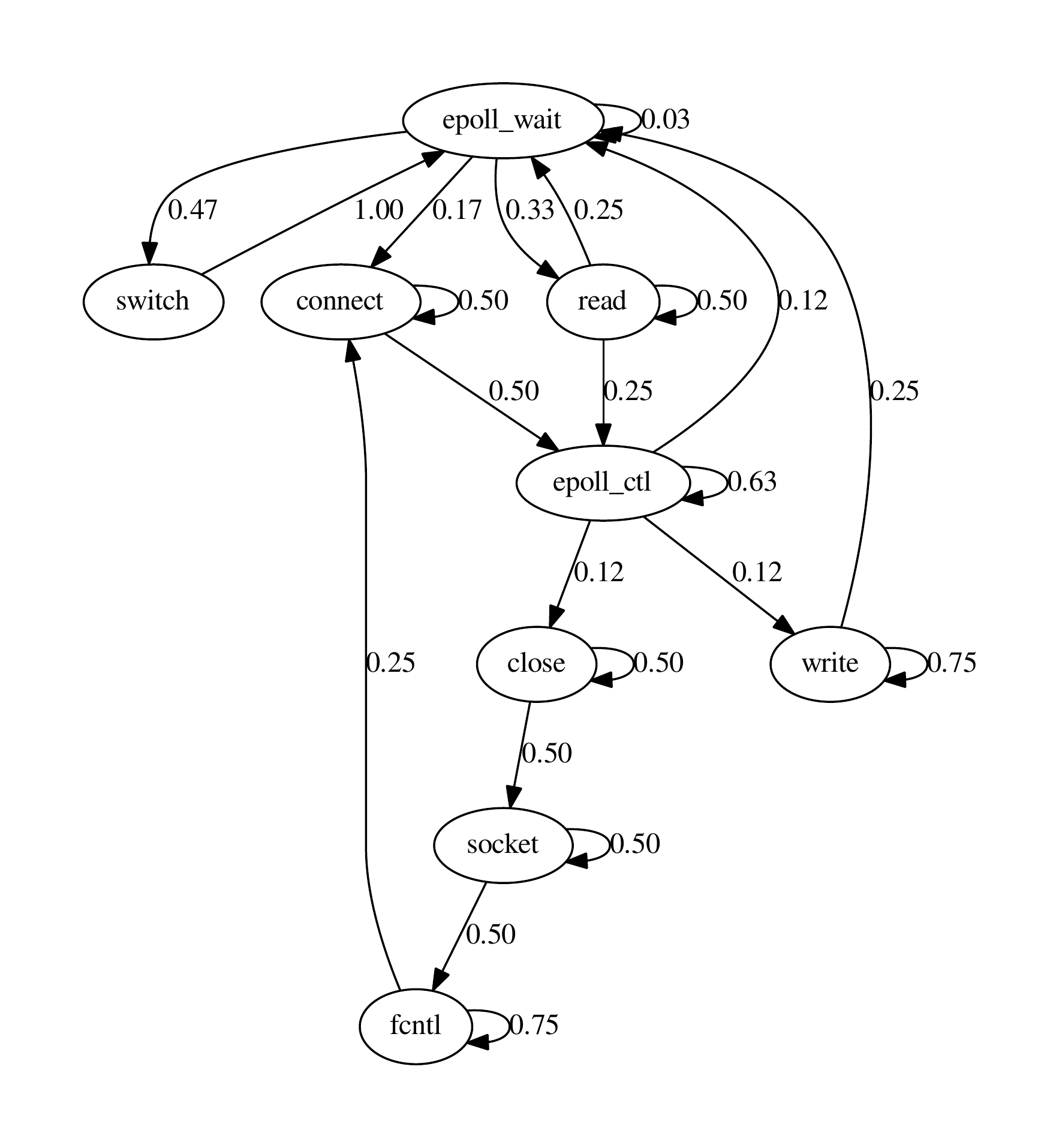}
    \includegraphics[width=0.3\textwidth]{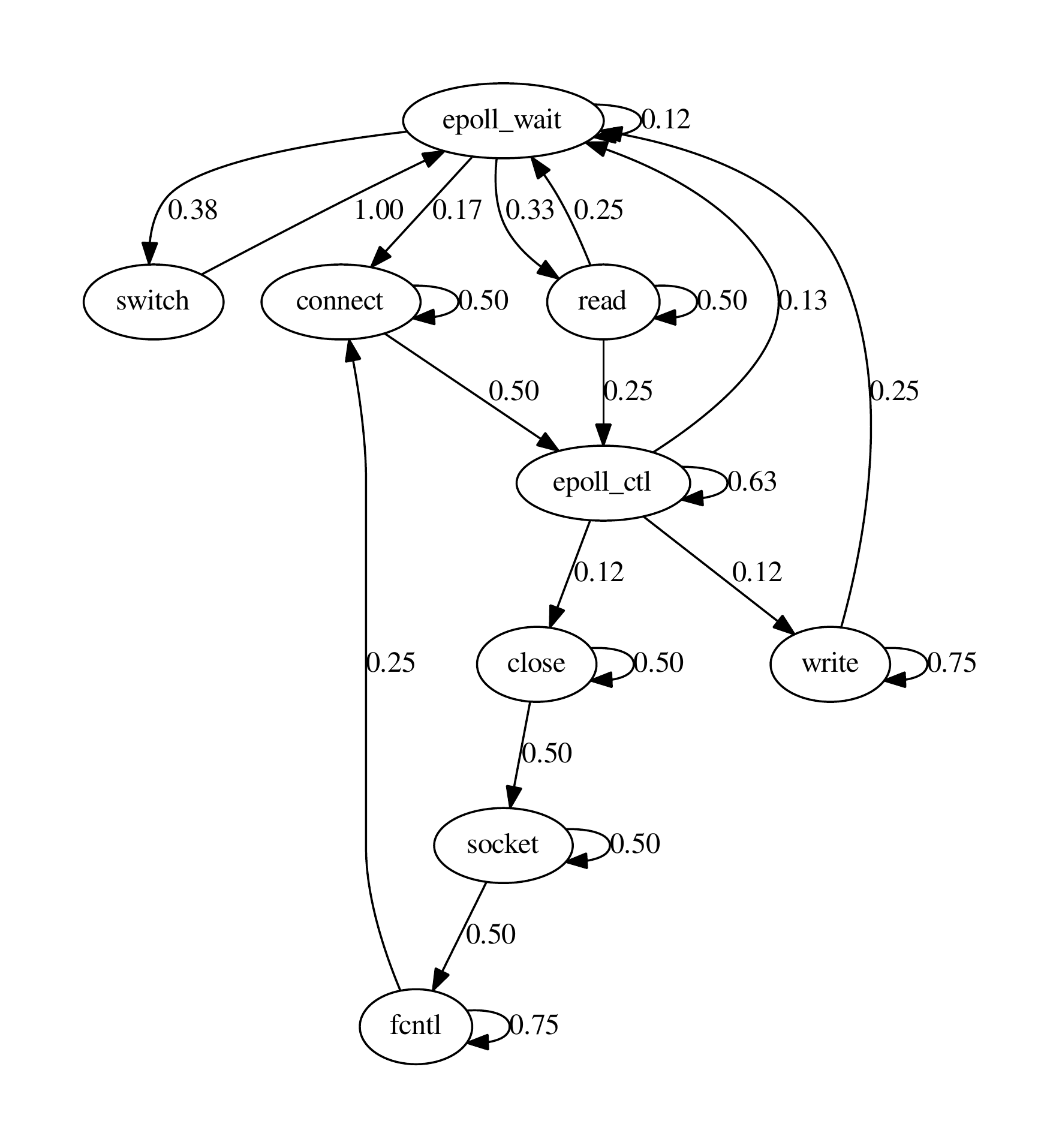}
    \caption{Client Markov chain for physical, e1000, and virtio. We dropped
    edges of weight less than $.001$ and renormalized.} \label{fig:markov}
\end{figure*}

Figure~\ref{fig:markov} shows pruned examples of these Markov chains for
ApacheBench across the testbeds. We combined 10 runs of each, dropped edges
with weight less then $.001$ and renormalized the probabilities around each
node. Client Markov chains had identical topology and nearly identical weights.
Server Markov chains disagreed on edge existence by up to 22\% and had more
weight variation on similar edges.

We can \emph{walk} a Markov chain using the sequence of system calls from a run
on a virtual or physical testbed to compute the probability the Markov chain
would generate that sequence. We show the walk/computation process with the
Markov chain from Figure~\ref{fig:Markov-eg} and the sequence (open, read,
read, read). We start at the node labeled ``open.''  The next element is the
first ``read,'' There is an edge from the ``open'' node to the ``read'' node
with weight $.99$, so the first pair of system calls (open, read) occurs with
probability $.99$. Similarly, there is an edge from the current node (``read'')
to a node with the next element of the sequence (the second read). This has
probability $.75$.  In Markov chains, each transition is independent, so the
probability of seeing the first part of the sequence (open, read, read) is $.99
\times .75$.  The next step is the same.  So the probability of the sequence
(open, read, read, read) is $.99 \times (.75)^2$.  The probability for the
sequence (open, read, read, close) is $.99 \times .75 \times .25$, since the
last transition goes from the ``read'' node to the ``close'' node, which
happens with probability $.25$  Because each additional sequence element
(system call) multiplies the final probability by a number less than 1, the
total can become quite small and sequence length becomes important for
comparison.  For example, for the Markov chain in Figure~\ref{fig:Markov-eg},
the most probable sequence of length 17 is open followed by 16 reads. This
sequence has lower probability than the sequence (open, error). Thus, we will
compare sequences of equal lengths, where the relative sizes of the
probabilities are a valid means of comparison, even though the absolute
probability for long sequences are tiny.

More formally, given a sequence of system calls and a Markov chain with
system-call node labels, we start at the Markov-chain node that matches the
first element of the sequence.  Moving to the next element in the sequence
(system call $s_i$) is a {\em transition} in the Markov chain to the node with
the label $s_i$. We continue the walk to the end of the sequence. The final
probability is the product of the weights of every edge we traverse with
multiplicities.

We built Markov chains using runs in the physical testbed by creating a node
for every system call in the sequence and an edge for each pair of consecutive
calls.  Edge $(c_i, c_j)$ is weighted by the percentage of times the call $c_i$
is followed by $c_j$. We then compared the relative probabilities when walking
the chain using a sequence from e1000 to that from virtio. For a baseline, we
also include the probability from walking the Markov chain with the physical
sequence. Since we ran ten iterations of each physical, e1000, and virtio, we
build ten Markov chains for each of the physical tests and compare to each of
the ten sequences of each type.

Consider a Markov chain created by sequence A.  When walking that chain using a
sequence B, is it possible that there is no edge for a particular transition
$(c_i, c_j)$.  That sequence did not appear in sequence A. We call this an {\em
invalid transition}, which gives the walk probability zero.

We apply this technique to the system-call sequences from ApacheBench with a
single thread. Table~\ref{tab:markov_walk_zeroes} shows how many times an
invalid transition occurs. We found that the invalid transitions typically
occur in the initialization phase of the workload so we also report the number
of invalid transitions when we skip the first 1 million (1M) system calls.

\begin{table}[ht]
    \centering
    \begin{tabular}{|r|r|r|}
        \hline
        Sequence & No Skip & Skip 1M \\
        \hline
        physical & 32 & 0 \\
        e1000    & 26 & 0 \\
        virtio   & 26 & 0 \\
        \hline
    \end{tabular}
    \caption{Invalid transitions from 100 walks of Markov chains using
    sequences from physical, e1000, and virtio.}
    \label{tab:markov_walk_zeroes}
\end{table}

Table~\ref{tab:markov_walk_values} shows the average probability when walking a
Markov chain built from the physical testbed with e1000 and virtio sequences
relative to the average when walking the same chain with physical sequences
(i.e. $P(\mbox{e1000})/P(\mbox{physical})$. We limit the sequences to 2M system
calls to ensure that we compare sequences of the same length (the shortest
sequence is just over 3M system calls). As before, we also report on the
probabilities when we skip the first 1M system calls.

\begin{table}[ht]
    \centering
    \begin{tabular}{|r|r|r|}
        \hline
        Sequence & No Skip & Skip 1M \\
        \hline
        % Relative probabilities, exponent only:
        e1000  & E+5531  & E+5402 \\
        virtio & E-33288 & E-28214 \\
        % Relative probabilities:
        % e1000  & 4.75E+5531 & 1.05E+5402 \\
        % virtio & 2.08E-33288 & 2.13E-28214 \\
        % Probabilities:
        %physical & 3.22E-696225 & 2.20E-696224 \\
        %e1000    & 1.53E-690693 & 2.30E-690822 \\
        %virtio   & 6.69E-729513 & 4.68E-724438 \\
        \hline
    \end{tabular}
    \caption{Relative probabilities from walking Markov chains built from
    physical testbed using sequences from e1000 and virtio to physical. We omit
    the bases as they are irrelevant given the magnitudes of the exponents.}
    \label{tab:markov_walk_values}
\end{table}

This chart shows that the e1000 sequences are many orders of magnitude more
probable than the physical sequences while the virtio sequences are many orders
less. We suspect that this is because we built the Markov chains with $N=1$
which does not include any system-call history. We performed some initial
experiments with $N=2$ and found that the physical sequences become more
likely, on average, than the e1000 sequences. We do not report on those results
further because the number of invalid transitions increases significantly so
that we have very few probabilities to average.

\section{Conclusion}
\label{sec:conclusion}

We have defined a first repeatable method to quantitatively compare physical
and virtual testbeds. We have applied these methods to simple network tests to
show our method's utility. Our experiments show that, for our simple workload,
our virtual testbed behaves reasonably close to its physical counterpart,
within our $10$\% threshold in many cases. We make this assessment using
multiple levels of abstraction from workload metrics to system interactions to
packet captures.

We hope this paper will encourage discussion and that other researchers will
extend these comparison methods for these simple tests and for more complex
tests.

\noindent\textbf{Acknowledgements} We thank Rob Johnson (VMWare), Ben Reed (San
Jose State University), and Jeff Boote (Netflix) for helpful discussions and
detailed comments on an early draft. Sandia National Laboratories is a
multimission laboratory managed and operated by National Technology and
Engineering Solutions of Sandia, LLC., a wholly owned subsidiary of Honeywell
International, Inc., for the U.S. Department of Energy's National Nuclear
Security Administration under contract DE-NA-0003525.

% include in citations even though we don't cite it explicitly
\nocite{Tange2011a}

{\footnotesize \bibliographystyle{acm}
\bibliography{main}}

\begin{thebibliography}{10}

\bibitem{apachebench}
ab -- apache http server benchmarking tool.
\newblock \url{https://httpd.apache.org/docs/2.4/programs/ab.html}.

\bibitem{minimega}
minimega: a distributed vm management tool.
\newblock \url{http://minimega.org/}.

\bibitem{owamp}
One-way ping (owamp).
\newblock \url{http://software.internet2.edu/owamp/}.

\bibitem{openvswitch}
Open vswitch.
\newblock \url{http://www.openvswitch.org/}.

\bibitem{sysdig}
sysdig.
\newblock \url{https://sysdig.com/}.

\bibitem{tcpdump}
tcpdump \& libpcap.
\newblock \url{http://www.tcpdump.org/}.

\bibitem{benzel2011science}
{\sc Benzel, T.}
\newblock The science of cyber security experimentation: the deter project.
\newblock In {\em Proceedings of the 27th Annual Computer Security Applications
  Conference\/} (2011), ACM, pp.~137--148.

\bibitem{chang1999network}
{\sc Chang, X.}
\newblock Network simulations with opnet.
\newblock In {\em Proceedings of the 31st conference on Winter simulation:
  Simulation---a bridge to the future-Volume 1\/} (1999), ACM, pp.~307--314.

\bibitem{Cheng:2016TransactionNetworking}
{\sc Cheng, L., Lau, F. C.~M., Cheng, L., and Lau, F. C.~M.}
\newblock Revisiting {TCP} congestion control in a virtual cluster environment.
\newblock {\em IEEE/ACM Transactions on Networking 24}, 4 (Aug. 2016).

\bibitem{Chun2003Planetlab}
{\sc Chun, B., Culler, D., Roscoe, T., Bavier, A., Peterson, L., Wawrzoniak,
  M., and Bowman, M.}
\newblock Planetlab: An overlay testbed for broad-coverage services.
\newblock {\em SIGCOMM Comput. Commun. Rev. 33}, 3 (July 2003).

\bibitem{forrest1996sense}
{\sc Forrest, S., Hofmeyr, S.~A., Somayaji, A., and Longstaff, T.~A.}
\newblock A sense of self for unix processes.
\newblock In {\em Proceedings 1996 IEEE Symposium on Security and Privacy\/}
  (May 1996), pp.~120--128.

\bibitem{Gamage:2013:TransCompSys}
{\sc Gamage, S., Kompella, R.~R., Xu, D., and Kangarlou, A.}
\newblock Protocol responsibility offloading to improve {TCP} throughput in
  virtualized environments.
\newblock {\em ACM Transactions on Computer Systems 31}, 3 (2013).

\bibitem{He:2016Sigcomm}
{\sc He, K., Rozner, E., Agarwal, K., Gu, Y.~J., Felter, W., Carter, J., and
  Akella, A.}
\newblock {AC/DC TCP}: Virtual congestion control enforcement for datacenter
  networks.
\newblock In {\em Proceedings of the 2016 ACM SIGCOMM Conference}, SIGCOMM '16.

\bibitem{hofmeyr1998intrusion}
{\sc Hofmeyr, S.~A., Forrest, S., and Somayaji, A.}
\newblock Intrusion detection using sequences of system calls.
\newblock {\em Journal of computer security 6}, 3 (1998), 151--180.

\bibitem{issariyakul2011introduction}
{\sc Issariyakul, T., and Hossain, E.}
\newblock {\em Introduction to network simulator NS2}.
\newblock Springer Science \& Business Media, 2011.

\bibitem{kivity2007kvm}
{\sc Kivity, A., Kamay, Y., Laor, D., Lublin, U., and Liguori, A.}
\newblock kvm: the linux virtual machine monitor.
\newblock In {\em Proceedings of the Linux symposium\/} (2007), vol.~1, Dttawa,
  Dntorio, Canada, pp.~225--230.

\bibitem{kroeger1999case}
{\sc Kroeger, T.~M., and Long, D.~D.}
\newblock The case for efficient file access pattern modeling.
\newblock In {\em Proceedings of the Seventh Workshop onHot Topics in Operating
  Systems, 1999.\/} (1999), IEEE, pp.~14--19.

\bibitem{lantz2010network}
{\sc Lantz, B., Heller, B., and McKeown, N.}
\newblock A network in a laptop: rapid prototyping for software-defined
  networks.
\newblock In {\em Proceedings of the 9th ACM SIGCOMM Workshop on Hot Topics in
  Networks\/} (2010), ACM, p.~19.

\bibitem{menon2005diagnosing}
{\sc Menon, A., Santos, J.~R., Turner, Y., Janakiraman, G.~J., and Zwaenepoel,
  W.}
\newblock Diagnosing performance overheads in the xen virtual machine
  environment.
\newblock In {\em Proceedings of the 1st ACM/USENIX international conference on
  Virtual execution environments\/} (2005), ACM, pp.~13--23.

\bibitem{minnich:2010eurosys}
{\sc Minnich, R., and Rudish, D.}
\newblock Ten million and one penguins, or, lessons learned from booting
  millions of virtual machines on hpc systems.
\newblock In {\em Proc. Workshop on System-level Virtualization for High
  Performance Computing in conjunction with EuroSys\/} (2010), vol.~10.

\bibitem{tcptrace}
{\sc Ostermann, S.}
\newblock tcptrace.
\newblock \url{http://www.tcptrace.org/}.

\bibitem{ricci2014Cloudlab}
{\sc Ricci, R., Eide, E., and Team, C.}
\newblock Introducing {CloudLab}: Scientific infrastructure for advancing cloud
  architectures and applications.
\newblock {\em ; login:: the magazine of USENIX \& SAGE 39}, 6 (2014), 36--38.

\bibitem{rizzo2013speeding}
{\sc Rizzo, L., Lettieri, G., and Maffione, V.}
\newblock Speeding up packet i/o in virtual machines.
\newblock In {\em Proceedings of the ninth ACM/IEEE symposium on Architectures
  for networking and communications systems\/} (2013), IEEE Press, pp.~47--58.

\bibitem{Tange2011a}
{\sc Tange, O.}
\newblock Gnu parallel - the command-line power tool.
\newblock {\em ;login: The USENIX Magazine 36}, 1 (Feb 2011), 42--47.

\bibitem{wang2010impact}
{\sc Wang, G., and Ng, T.~E.}
\newblock The impact of virtualization on network performance of amazon ec2
  data center.
\newblock In {\em INFOCOM, 2010 Proceedings IEEE\/} (2010), IEEE, pp.~1--9.

\bibitem{warrender1999detecting}
{\sc Warrender, C., Forrest, S., and Pearlmutter, B.}
\newblock Detecting intrusions using system calls: alternative data models.
\newblock In {\em Proceedings of the 1999 IEEE Symposium on Security and
  Privacy (Cat. No.99CB36344)\/} (1999), pp.~133--145.

\bibitem{whiteaker2011explaining}
{\sc Whiteaker, J., Schneider, F., and Teixeira, R.}
\newblock Explaining packet delays under virtualization.
\newblock {\em ACM SIGCOMM Computer Communication Review 41}, 1 (2011), 38--44.

\end{thebibliography}

%\theendnotes

\end{document}